\documentstyle[aas2pp4]{article}

\newcommand{\sci}[2]{\mbox{$ #1 \times 10^{ #2 }$}}   
\newcommand{\F}[1]{Figure~\protect\ref{#1}}   
\newcommand{\Fdot}[1]{Fig.~\protect\ref{#1}}  
\newcommand{\T}[1]{Table~\protect\ref{#1}}    
\newcommand{\R}{\reference}
\newcommand{\eg}{e.g.,\ }
\newcommand{\ea}{et~al.}
\newcommand{\ie}{i.e.,\ }

\journalid{}{}
\articleid{}{}
\lefthead{Walker, Mihos, and Hernquist}
\righthead{Sinking Satellites}
\slugcomment{Submitted to the {\it Astrophysical Journal}}

\begin{document}

\title{Quantifying the Fragility of Galactic Disks in Minor Mergers}

\author{Ian R. Walker,
        J. Christopher Mihos\altaffilmark{1,2},
        \and Lars Hernquist\altaffilmark{3}}
\affil{Board of Studies in Astronomy and Astrophysics,\\
	 University of California, Santa Cruz, CA 95064\\
	 {\it iwalker, hos, lars\\
              @ucolick.org}}

\altaffiltext{1}{Hubble Fellow}
\altaffiltext{2}{present address:  Johns Hopkins University, Department of
                 Physics and Astronomy, Baltimore, MD 21218}
\altaffiltext{3}{Alfred P. Sloan Foundation Fellow, Presidential Faculty
Fellow}

\begin{abstract}
We perform fully self-consistent stellar dynamical simulations of the accretion
of a companion (``satellite'') galaxy by a large disk galaxy to investigate the
interaction between the disk, halo, and satellite components of the system
during a merger.  Our fiducial encounter begins with a satellite in a prograde,
circular orbit inclined thirty degrees with respect to the disk plane at a
galactocentric distance of six disk scalelengths.  The satellite's mass is 10\%
of the disk's mass and its half-mass radius is about 1.3~kpc.  The system is
modelled with 500\,000 particles, sufficient to mitigate numerical relaxation
noise over the merging time.  The satellite sinks in only $\sim 1$~Gyr and a
core containing $\sim 45\%$ of its initial mass reaches the centre of the disk.
With so much of the satellite's mass remaining intact, the disk sustains
significant damage as the satellite passes through.  At the solar circle we
find that the disk thickens $\sim 60\%$, the velocity dispersions increase by
$\Delta\mbox{\boldmath$\sigma$} \simeq (10,8,8)$~km/s to
$(\sigma_R, \sigma_\phi, \sigma_z) \simeq (48, 42, 38)$~km/s, and the
asymmetric
drift is unchanged at $\sim 18$~km/s.  Although the disk is not destroyed by
these events (hence ``minor'' mergers), its final state resembles a disk galaxy
of earlier Hubble type than its initial state, thicker and hotter, with the
satellite's core enhancing the bulge.  Thus minor mergers continue to be a
promising mechanism for driving galaxy evolution.
\end{abstract}


\clearpage

\section{Introduction}

The business of simulating interactions between galaxies originated with
Holmberg (1941!) who found that a close encounter could raise impressive and
observable tidal distortions and even thermalize enough orbital energy to
result
in capture.  Although he was studying the clustering of galaxies, this work
came
long before the recognition of large-scale structure and it was revisited only
rarely (\eg by Pfleiderer \& Siedentopf~(1961); Pfleiderer~(1963)) because
of the cherished view of galaxies as ``island universes'' which rarely
interact.
However, a number of observed galaxies show signs of interaction
(Arp~1966) and the subject was reintroduced around 1970, particularly by
Toomre \& Toomre (1972), who successfully modelled several peculiar galaxies
with collisions.  The expected frequency of encounters grew with the belief
that galaxies are embedded in extensive, massive, dark halos which increase
cross-sections for major collisions and enable the orbits of satellite galaxies
to decay by dynamical friction (Tremaine~1981).  Now, more than fifty years
after Holmberg's illuminating efforts, the prevalence of the hierarchical
clustering picture for the formation and evolution of structure makes the study
of galaxy interactions a very active field and a wide range of phenomena are
thought to be linked to one variety of galaxy interaction or another.

Major mergers between spirals of comparable mass have received the
most attention.  The intermediate stages of such collisions explain some
of the most spectacular objects observed, reproducing their messy profiles
and extended tidal tails.  Because the stellar component of the remnant relaxes
toward a de~Vaucouleurs profile, such mergers appear to drive evolution along
the Hubble sequence toward elliptical galaxies (Negroponte \& White~1983;
Barnes~1988, 1992; Hernquist~1992, 1993a).  They may also trigger starbursts
or nuclear activity when gas is driven to the remnant's centre as the
progenitors coalesce (Mihos \& Hernquist~1994a; Barnes \& Hernquist~1995).

Less spectacular but more common, mergers in which a disk galaxy accretes a
smaller object (``minor mergers'') have analogous effects, namely, the
stirring-up of disks and the generation of peculiar features (Quinn \&
Goodman~1986; Quinn, Hernquist \& Fullagar~1993), bulge-building (a smaller
step along the Hubble sequence toward Sa/S0 (Schweizer~1990)), and enhanced
star formation or nuclear activity (Hernquist~1989; Mihos \& Hernquist~1994b).
Specifically, the early stages of such interactions (or even grazing
encounters)
can induce spiral arms, bars, warps, or ``bridges'' and full merging can
thicken
and dynamically heat disks.  The tidal stripping of a satellite can produce
features such as long tails or counterstreaming groups and its accretion
introduces a new stellar population into the disk.  A galactic bulge may be
formed or enlarged by any combination of (a) the satellite's core surviving to
the centre of the disk, (b) gas being driven to the centre and converted to
stars, or (c) disk stars being stirred up out of the disk plane.  Combine this
bulge-building with the smoothing of disk features which results from the
increased velocity dispersion and the primary galaxy looks more like an S0 than
it did before.  For a summary of these ideas see the review by Barnes \&
Hernquist (1992) or any one of a number of conference proceedings
(that of Wielen (1990) is comprehensive).

Those classic, messy objects which spark observational interest in major
mergers have been known for decades.  Interest in minor mergers has mostly
developed more recently because the subtler features attract less attention
and can be much harder to detect and study.  However, considerable
observational
data has accumulated.  The Milky Way has a number of satellite companions,
as do most large galaxies.  There is evidence that some quasar activity could
be associated with minor mergers (Bahcall, Kirhakos, \& Schneider~1995).  Grand
design spiral structure is sometimes associated with small, close companions:
M51 is a classic example.  A large fraction of disk galaxies are warped
(Binney~1992).  ``Thick disks'', hot, flattened components with kiloparsec
scaleheights, have been discovered in our Galaxy (Gilmore \& Reid~1983;
Gilmore, Wyse, \& Kuijken~1989) and others (Burstein~1979; van der Kruit \&
Searle~1981).  The Magellanic Clouds and Stream have been known for a long
time,
but the discovery of the close, tidally stretched Sagittarius dwarf galaxy
(Ibata, Gilmore, \& Irwin~1994) and the recognition of a distinct, possibly
accreted population of blue, metal-poor stars in the solar neighbourhood
(Preston, Beers, \& Shectman~1994) have generated recent excitement.  Most
recently, we have found that the bulge of Hickson 87a very closely matches the
bulge produced in one of our simulations (Mihos \ea~1995).  The parallel
between
the observational results listed here and the minor merger effects listed in
the previous paragraph drives the current interest in discrete accretion
events.

Tremaine (1981) used the mean radial distribution of satellite companions of
large galaxies and sinking rates derived from dynamical friction in an
isothermal halo to estimate the total mass accreted by a large galaxy over a
Hubble time.  He found that a typical large galaxy can consume 20\%--30\% of
its
own mass, very roughly, motivating several attempts to quantify the resilience
of cold, thin disks to minor mergers:  What {\em are\/} the implications of the
chunk-wise accretion of so much mass?  Quinn \& Goodman (1986) measured the
effect of the disk on these sinking rates, studying the dynamics of the
disk-satellite interaction with analytic and restricted $N$-body techniques.
They noted strong heating of their simulated disks.  Quinn, Hernquist, \&
Fullagar (1993) used $N$-body experiments to survey orbital parameters,
focussing on such observable disk properties as density structure and
kinematics.  They found that single accretions of about 10\% of the disk's mass
could, in the solar neighbourhood, double the disk's thickness and expand the
velocity ellipsoid by fifty per cent (more in the radial direction), producing
something similar to the Milky Way's thick disk.  The analytic work of T\'{o}th
\& Ostriker (1992) also found disks to be quite fragile.  They turned
Tremaine's
question around by using the observed abundance of undisturbed, cold, thin disk
galaxies to limit the mean total mass accreted and, by inference, $\Omega_0$.
They concluded that the Milky Way cannot have accreted more than 4\% of the
mass
inside the solar circle in the last 5~Gyr.  Since $\Omega=1$ models predict
that accretion continues at late times, they argued that $\Omega_0<1$.

Successes notwithstanding, these studies had the drawback of not being fully
self-consistent, modelling the dark matter halo and sometimes also the
satellite
as rigid bodies.  Modelling a component with interacting particles (thus making
it responsive, or ``live'') increases both computational expense and noise and
might seem frivolous for the halo which is, after all, invisible.  However,
rigid representations lack internal degrees of freedom and thus the disk was
the {\em only\/} sink for the satellite's energy in earlier investigations.
These studies were in some sense measures of the maximum damage that could be
done to the disk, not of the typical response.  Also, a live halo is able to
propagate disturbances into the disk (Weinberg~1995a), a pathway which is cut
off when a rigid halo is used.

Confidence in the results of Quinn \ea~(1993) is undermined by several other
technical limitations and compromises, apart from the use of a rigid halo.
Their simulations used 32\,768 disk particles and 4096 satellite particles and
were thus fairly noisy.  Their satellite was initialized with some fraction of
its mass not bound because the tidal radius was not taken into account.  This
mass was left behind at large radii and did not participate in the stirring of
the disk.  Also, their disk was abruptly truncated at the initial radius of the
satellite's orbit, interfering with the disk-satellite coupling.  This allowed
the satellite to orbit several times at large radii before sinking, making
sinking times unreliable but also allowing the satellite to shed even more mass
at large radii, further reducing its potency for scattering disk particles.

Advances in computer technology now permit the use of an order of magnitude
more particles than used by Quinn \ea~(1993).  Here we attempt to remedy the
shortcomings of earlier studies with fully self-consistent N-body simulations
large enough to eliminate numerical relaxation noise over the merging time
while including the dynamical interplay of all the components of the system.

\section{Models and Methods}

\subsection{Galaxy Models}

Our models consist of exponential disks, truncated isothermal halos,
and Hernquist model satellites (Hernquist~1990a) in the mass ratio 10:58:1.
The disk and halo each contain 45\% of the simulation particles while the
satellite contains the remaining 10\%, an arrangement which provides better
sampling of the luminous components.  The specific profiles used are:
\begin{eqnarray}
   \rho_{\mbox{disk}}(R,z)& = & \rho(0)e^{-R/h}\cosh^{-2}(z/z_0),\label{eq:
d}\\
   \rho_{\mbox{halo}}(r)  & = & \rho(0)\frac{e^{-r^2/r_c^2}}{1
+r^2/\gamma^2},\\
   \rho_{\mbox{sat}}(r)   & = & \frac{Ma}{2\pi r(r+a)^3}.
\end{eqnarray}
The satellite is truncated at its initial tidal radius with respect
to the primary.  Parameter values are listed in \T{params}.

Velocities are initialized from moments of the collisionless Boltzmann equation
in a procedure described by Hernquist (1993b).  For each dark matter particle,
the velocity ellipsoid suitable for its location is calculated from the moment
equations and then its velocity components are randomly selected from gaussians
corresponding to that ellipsoid.  For disk particles, velocities are
constrained
to make the radial dispersion proportional to the square root of surface
density
($\sigma_R \propto e^{-R/2h}$; Freeman 1993) and are normalized to a chosen
value of $Q(R_{\odot})$.  The vertical dispersion is given by $\sigma_z^2 =
\pi G \Sigma(R) z_0$ (the so-called isothermal sheet).  The azimuthal
dispersion
is given by the epicycle approximation ($\sigma_{\phi}^2 =
\sigma_R^2 \kappa^2/(4\Omega^2)$) and the azimuthal streaming velocity by the
cylindrical moment equations.  The initial disk structure is shown in \F{init}.

This approach is not without its drawbacks since it only approximates a
distribution function and thus does not initialize the particles in a true
equilibrium configuration.  When this initial state is allowed to evolve in
isolation (\ie without a satellite) it rapidly shifts to an equilibrium
configuration whose subsequent evolution is governed only by numerical
relaxation (simulation noise).  However, this shift can be large.  \F{initcomp}
illustrates this by superposing the initial disk structure on the structure
after several time steps.  Note in particular that the central dispersions
have dropped by about twenty per cent.

For our purposes this technique is adequate because the galaxy is about to be
stirred up by an infalling neighbour.  As long as we are careful to measure
``heating'' and ``thickening'' relative to the coeval isolated disk galaxy
rather than the initial state, our analysis will be sound.  On the other hand,
studies involving isolated disk galaxies (\eg instability studies) would
require
better initial conditions, both for tighter control of initial structure and
less disturbance from transients.  Improvements can be obtained from
higher order moments but for future work the semi-analytic distribution
functions of Kuijken \& Dubinski (1995) may offer a more direct approach.

\subsection{Encounter}

We choose to focus on an encounter in which the satellite starts on a circular,
prograde orbit with a radius of six disk scalelengths inclined 30\arcdeg\ with
respect to the disk plane.  This encounter was found by Quinn \ea~(1993)
to illustrate most
of the important phenomena.  Note that the satellite is not grown adiabatically
in this orbit; it is simply switched on at $t=0$.  Although the initialization
procedure tries to account for the effect on the satellite by truncating it at
its tidal radius, the disk is thrown somewhat out of equilibrium by the sudden
change in potential at its edge.  However, as described above and demonstrated
in \F{initcomp}, the disk is initially out of equilibrium anyway, and the
transient behaviour within $R=15$~kpc is identical to that for the isolated
galaxy shown in \F{initcomp}.  Nevertheless, as computers become ever faster,
it will be prudent to invest some cpu time in moving the starting radius out
farther for the sake of more realistic orbital evolution, tidal stripping,
flaring and warping of the disk, et cetera.

For the sake of discussion, units are scaled to the Milky Way such that the
disk has scalelength $h=3.5$~kpc and mass $M_d = \sci{5.6}{10} M_{\odot}$
(Bahcall, Schmidt, \& Soneira 1983). Other parameters are given in \T{params}.
The evolution was followed with a treecode (Barnes \& Hut~1986;
Hernquist~1987,~1990b) using terms up to quadrupole order,
a tolerance parameter (``opening angle'') $\theta =0.7$, and a timestep
$\Delta t \simeq 2$~Myr.  Energy was conserved to 0.1\% over 1.2~Gyr and to
0.2\% over 2.5~Gyr.  Simulations which include hydrodynamics have also been
studied and are reported separately (Hernquist \& Mihos 1995).

\begin{table}[ht]
   \begin{tabular}{lc}
      \tableline
      \tableline
      Quantity                         &        Value              \\
\tableline
      Disk                             &                           \\
      \ \ Mass $M_d$                   & \sci{5.6}{10} $M_{\odot}$ \\
      \ \ Scalelength $h$              &       3.5 kpc             \\
      \ \ Scaleheight $z_0$            &       700 pc              \\
      \ \ Softening $\epsilon_d$       &       140 pc              \\
      \ \ Toomre $Q(R_{\odot})$        &        1.5                \\
      \ \ $N_d$                        & $0.45N$\tablenotemark{a}  \\
      Halo                             &                           \\
      \ \ Mass $M_h$                   & \sci{3.25}{11} $M_{\odot}$\\
      \ \ Scalelength $\gamma$         &       3.5 kpc             \\
      \ \ Cutoff Scale $r_c$           &       35 kpc              \\
      \ \ Softening $\epsilon_h$       &       700 pc              \\
      \ \ $N_h$                        & $0.45N$\tablenotemark{a}  \\
      Satellite                        &                           \\
      \ \ Mass $M_s$                   & \sci{5.6}{9} $M_{\odot}$  \\
      \ \ Characteristic Radius $a$\tablenotemark{b}  &   525 pc   \\
      \ \ Softening $\epsilon_s$       &        70 pc              \\
      \ \ $N_s$                        &  $0.1N$\tablenotemark{a}  \\
      \ \ Orbit $R_{\mbox{init}}$      &       21 kpc              \\
      \tableline
   \end{tabular}
   \caption{Simulation Parameters}  \label{params}
   \tablenotetext{\rm a}{$N=500\,000$ particles for the largest simulation.}
   \tablenotetext{\rm b}{The half-mass radius is $(1 + \sqrt{2})a$.}
\end{table}

\subsection{Quality Control}     \label{sect: qualcon}

Modelling a halo (or any other component) with particles does have a
disadvantage in that the potential is less smooth than a rigid, analytic halo's
and scattering is thus enhanced.  Compared with real galaxies, simulated
galaxies have a relatively small number of relatively massive particles whose
wide, deep potential wells are able to deflect one another even if they pass
at a significant distance.  This excess scattering, called 2-body relaxation
noise, is to some extent suppressed by ``softening'', artificially reducing the
force between close particles to prevent strong collisions.  However, random
clumping (``shot noise'') creates potential fluctuations on all scales and
softening does not help with scattering off large clumps.  The contrast between
random clumps and the background is reduced as more particles are used; if we
are to detect small but legitimate structural and kinematic effects in our
simulations, we must reduce the excess scattering as much as possible.
To quantify this, we ran several simulations of the disk galaxy in isolation
(\ie without the satellite).  \F{isostruct} shows the evolution of the disk
thickness and velocity ellipsoid at the solar radius in simulations with
$N=45\,000$, which is roughly the size of the Quinn \ea~(1993) simulations,
$N=90\,000$, $N=225\,000$, and $N=450\,000$.  The largest run experienced very
little change over the sinking time ($\sim 1$~Gyr) so we expect any evolution
we observe in our satellite encounters to be signal rather than noise.

Shot noise has also been observed to have a global manifestation.  Disk models
with realistic profiles and velocity dispersions tend to be quite lively in
that
they amplify density perturbations which can then feed into instabilities and
produce spiral or bar features (Toomre~1981; Binney \& Tremaine~1987;
Sellwood~1989; Hernquist~1993b; Weinberg~1995b).  This is exacerbated in models
with live halos because clumps of massive particles create wakes in the disk.
Since the halo dominates the potential and is the most poorly-sampled
component,
it is the principal source of shot noise.  This is apparent in \F{isomodes},
which shows the growth of the bar ($m=2$) mode in our isolated galaxies and in
a run with 45\,000 disk particles and a rigid (noiseless) halo whose bar mode
seems not to grow at all.  Because the contrast between the random clumps and
the underlying smooth distribution is smaller when $N$ is larger, the
instability is seeded at a lower amplitude and thus sets in at later times in
larger simulations.  Again, it is apparent that noise has been suppressed for
the relevant timescale in the largest simulation and that any significant bar
growth in our satellite encounter can be assumed to be induced by the
satellite.

Thus we use the 450\,000-particle disk galaxy with a 50\,000-particle
satellite to model our chosen encounter.  \F{isostruct} and \F{isomodes}
demonstrate the need for large $N$ to beat down noise over long timescales:
\case{1}{2}-million particles is only adequate for about two billion years.
Many more halo particles must be used for examining phenomena with timescales
longer than a few gigayears, such as retrograde encounters.

\section{Results}                \label{sect: results}

The basic sequence of events is illustrated in \F{snaps} which shows both
face-on and edge-on views of the disk and satellite at regular intervals of
about 125~Myr.  The satellite loses most of its vertical motion while
completing
only $\sim 1.5$ orbits, settling into an orbit coplanar with the disk.
This orbit then decays quite rapidly.  The satellite sheds mass all along
its orbit but its core survives and arrives at the centre of the disk.

In this paper we discuss the decay of the satellite orbit and the satellite's
effect on the disk.  However, the disturbance continues to evolve for some
time after the merger is complete.  The evolution of the central regions at
late times is presented in a separate paper (Mihos \ea~1995), but here we
give a brief description.  \F{snaps} illustrates the disk's global response to
the satellite.  After the merger (\Fdot{snaps}, last frame), full axisymmetry
is not restored:  a bar has been induced and it drives further evolution in the
disk's inner two scalelengths.  Note that this bar is induced by the satellite,
not by amplified noise.  Because the bar is unstable to vertical bending
(Raha \ea~1991), it flexes and eventually kicks material out of the disk plane,
generating a small, flattened bulge with an X-morphology (Mihos \ea~1995).
For our purposes, however, it is sufficient to note that, apart from changes
associated with the bar, the disk structure does not change much, postmerger.

\subsection{Orbital Decay}       \label{sect: orbit}

\F{orbit} shows various aspects of the orbital decay.  The upper left panel
shows {\em cylindrical\/} radius versus time.  This quantity does not decrease
monotonically because the satellite is initially orbiting in an inclined plane.
Once the satellite has settled closer to the disk plane, there is a knee in the
$R$\/~versus~$t$\/ curve and the orbit decays very rapidly.  The upper right
panel shows the orbit from the North galactic pole and reveals that the
satellite falls most of the way to the centre in only two orbits
($R\sim 5$~kpc).  The lower left panel shows the altitude versus time and the
lower right shows altitude versus radius, demonstrating that the satellite
settles most of the way to the disk plane while still at large radii.

In an attempt to quantify the importance of the halo compared with the disk
over
the course of the encounter we calculate the torque on the satellite due to the
disk and halo separately (\Fdot{torques}, upper panel). The two torques are
comparable initially but the disk dominates overall.  A rough integration of
the curves in the upper panel of \F{torques} shows that $\sim 75\%$ of the
total time-integrated torque comes from the disk.

This is attributable to a resonant enhancement of dynamical friction which
occurs because a satellite on a prograde orbit moves with the nearby disk
particles and thus interacts strongly with them.  The interaction dynamics were
discussed in detail by Quinn \& Goodman~(1986), but basically leading disk
particles are pulled back by the satellite so that they lose angular momentum
and drop into lower orbits, clearing a gap in the disk in front of the
satellite
(\Fdot{snaps}).  Trailing particles gain angular momentum and move into higher,
slower orbits and thus do not overtake the satellite.  Once this configuration
develops, with the density higher behind the satellite than before it, the disk
torque increases dramatically and the satellite quickly sinks to the centre.
The lower panel of \F{torques} superposes the bar mode strength, the disk
torque, and the orbital decay to show that these coincide.  Note that this
mechanism transfers most of the satellite's orbital energy into the potential
energy of the disk.  If, when the satellite pulls a trailing disk particle
forward, promoting it to a higher orbit, the particle were to move  from one
circular orbit to another and the rotation curve were exactly flat, then the
particle's kinetic energy would not change a bit and all the energy flowing
from satellite to disk would appear as potential energy.

Clearly, the process is aided by the settling of the satellite toward the disk
plane, although the rapid radial decay of the orbit actually begins before the
vertical motion has damped out completely (\Fdot{orbit}).  The process also
relies on the approximate circularity of the orbit in order for the velocity of
the satellite to match the velocities of the disk particles. The global nature
of the response presumably requires an approximately flat rotation curve to
make
the resonant region extensive. The strength may depend on the ease of
excitation
of the bar mode; however, a run with a flattened Hernquist model bulge
(semiaxes
$a=0.7$, $c=0.35$~kpc) yielded almost identical sinking time and heating.

The halo, whose particles do not collectively orbit with the satellite, is
unable to undergo strong collective interaction and so cannot contribute such
a large torque.  Of course, in real life the satellite does not start at the
edge of the disk.  The torque on a satellite at large distance will be
dominated
by the halo and the satellite will take a long time to sink toward the disk
where the process described above can set in.  In fact, the disk's contribution
to the torque in the early stages of the decay of a prograde orbit can oppose
the halo's (Quinn \& Goodman~1986):  the satellite can gain angular momentum
from the disk in a manner analogous to the Moon's gain of angular momentum from
the tidal bulge it raises on the Earth.  This makes the truncation of the disk
by Quinn \ea~(1993) problematic.  Their Figure~11 shows orbital evolution
similar to that seen here but delayed.  The knee in their $R$\/~versus~$t$\/
curve comes at $\sim 2$~Gyr and the total sinking time is $\sim 2.5$~Gyr.

We test our simulations for convergence by running four simulations of our
encounter with 50\,000, 100\,000, 250\,000, and 500\,000 particles,
respectively.  The upper panel of \F{decay} shows the orbital decay curves.
They converge well and, since our simulations are fully self-consistent, we
conclude that we finally have reliable estimates for the sinking time.  It is
not very long: at 1~Gyr, it is only $\sim 10\%$ the age of the Milky Way's
disk.
(For comparison, we overlay the result from a simulation identical to our
largest in every respect but for a rigid halo.  The sinking time is some
50\%--80\% longer (though still much shorter than in Quinn \ea~(1993)) which
testifies to the importance of a self-consistent treatment.)  Even more
dramatic
is a coplanar ($i=0\arcdeg$), prograde encounter in which the satellite sinks
from six scalelengths in 0.6~Gyr (\Fdot{decay}, lower panel).  Evidently, once
a
satellite arrives in the disk on a prograde orbit, it has only a short time to
live.  This is in stark contrast to retrograde and polar orbits.  We tried
evolving them for $\sim 3\case{1}{4}$~Gyr and what little was left of the
satellite core had just reached the galactic centre (\Fdot{decay}, lower
panel).
As is apparent from Figures~\ref{isostruct} and \ref{isomodes}, noise becomes a
problem on these timescales.

\subsection{Disk Structure}      \label{sect: structure}

\F{structure} shows the disk thickness and velocity dispersions as functions of
cylindrical radius.  The solid lines show the structure at $\sim 1.2$~Gyr
(which allows $\sim 200$~Myr for things to settle after the satellite reaches
the centre) and the dashed lines show the structure of the coeval isolated
galaxy.  The entire disk thickens with respect to the isolated galaxy but much
more so at large radii where the satellite still had significant vertical
motion: more than 200\% thickening is seen beyond $R \sim 15$~kpc, as opposed
to
$\sim 50\%$ at $R \sim 5$~kpc and only $\sim 10\%$ at the centre.  In contrast,
the disk thickens much more uniformly in the coplanar encounter but only by
about 20\%, or 100~parsecs.  The velocity dispersions in \F{structure} increase
at least 10~km/s at all radii.  At the centre, $\sigma_R$ increases by 50~km/s
and $\sigma_z$ increases by 30~km/s, which is interesting because the central
thickness is essentially unchanged.  This reflects the increased depth of the
central potential due to the satellite core (see \S~\ref{sect: satrem}).
Note that the solid lines in \F{structure} are more extended than the dashed:
the disk spreads radially as well as vertically, storing much of the energy
it gains as potential energy.

The evolution of these quantities at the solar circle (8.0 kpc) is shown in
\F{solar}.  Recall that these quantities are azimuthally averaged and that
the disk is not axisymmetric at intermediate times:  the numbers calculated
for these times should not be trusted absolutely.  However, trends in the
disk's
behaviour are more reliable.  The figure shows that the satellite's effect is
felt in the radial dispersion first and in the other quantities only when the
satellite reaches the solar circle.  This pattern occurs in all the simulations
and so should be robust.  The increase in the radial dispersion is due to
azimuthal averaging when there is radial streaming in the bar.  It is thus not
surprising that $\sigma_R$ increases early because the bar develops before the
satellite reaches the solar circle. The asymmetric drift also responds to the
passage of the satellite but settles back to about 18~km/s so that, within the
noise, it has hardly changed at all.  The other quantities settle but not to
their original values.  In the end, the disk thickens by about 60\%.  The
radial
dispersion goes from about 35 km/s to about 48 km/s, the azimuthal from about
32 to about 42, and the vertical from 28 to 38. Compare these final values with
those from the isolated disk: around $t\sim 1.2$~Gyr, it has actually shrunk
about 3\%, heated to (38, 34, 30)~km/s, and has an asymmetric drift of 16~km/s.
The net heating attributable to the action of the satellite is thus
(10, 8, 8)~km/s.

The satellite acts to heat and thicken the disk by scattering disk particles.
Because the satellite is concentrated and moving, disk particles on close
trajectories can encounter slightly different parts of the satellite potential
and can be deflected in slightly different directions.  This creates a greater
variety of local trajectories, \ie the velocity dispersion increases.  So,
while the satellite's energy is absorbed into the potential energy of the disk
(and the stripped satellite material), the heating and thickening of the disk
occur more by pure scattering, the conversion of disk orbital energy to disk
thermal energy.  \F{solar} supports this view by indicating that heating and
thickening of the disk at the solar radius occur as the satellite passes
through
that radius and are thus essentially local processes.

\subsection{Satellite Remnant}   \label{sect: satrem}

That the satellite's core arrives as a distinct lump in the galactic centre
is apparent from the integrated mass distribution and surface density of the
satellite remnant (\Fdot{satstruct}).  About 45\% of the satellite's mass lies
within 2~kpc of the centre at the end, enhancing the peak surface density by a
factor of about 2.5 with respect to the disk.  Contrast this with the
retrograde
case in which the satellite barely survives.

Estimating mass loss with a simple density criterion based on the ``tidal
radius'' for a body in orbit about another (Binney \& Tremaine 1987) is
inadequate for the process we are studying.  Obviously, any treatment which
approximates the disk as a point mass when the satellite is orbiting within the
disk is inappropriate, but such estimates have been used (\eg by T\'{o}th \&
Ostriker (1992)) to relate mass loss to orbital radius.  To see the potential
for error, consider that the density ratio criterion does not make use of the
spin of either body, hence cannot distinguish between prograde and retrograde
events, and thus predicts the same mass loss for both.  Although our retrograde
simulation is unreliable for its fine structure because of the growth of noise
over the long sinking time, its bulk properties are sound.  The initial
configuration is identical to that of our main simulation except for the sign
of
the orbital angular momentum; the mass loss history is quite different.  By the
end of the simulation, the satellite core has been pared down to only 10\% of
the initial satellite mass (\Fdot{satstruct}, upper panel).  If half the
satellite survives to the centre in the prograde case while the satellite is
almost entirely disrupted in the retrograde case then a mass loss estimator
which does not distinguish these cases is all but useless.  Since it takes into
account neither the angular momentum of the objects nor the timescale of
interaction, the density criterion does not capture the physics involved and is
misapplied in this problem.  Given that a satellite's mass loss determines its
potency for scattering disk particles, any study relying on simple analytic
estimates of the mass loss must be interpreted very carefully.

As for the material which is stripped from the satellite, it is distributed in
a thickened, flared disk, similar to the final distribution of disk material
but thicker.  Because of its low surface density, this material would not stand
out in an external galaxy; only the core, visible as a central brightness
enhancement or small bulge (the peak in the dotted curve in the second panel
of \F{satstruct}), would be conspicuous.   We note here that Hickson 87a,
a galaxy whose luminosity structure matches our model quite closely, has
such a peak (Mihos \ea~1995).  \F{combstruct} shows the
structure of the satellite remnant and disk combined, along with the disk and
satellite separately for comparison.  Here again the core stands out as a
distinct component.  Otherwise, the satellite remnant is structurally and
kinematically like a thick, flared, hot disk.  There is little difference
between the disk structure and the combined structure (except at very large
radii) because the surface density of the remnant is just too small.  Near the
solar circle, the satellite remnant is somewhat thicker and hotter than the
disk material but it does not lag in its rotation (within the noise).

Since the satellite material does not alter the combined structure much,
especially at $R_{\odot}$, it is not clear that it could be distinguished
without a spectral signature (\eg the blue, metal-poor population discussed by
Preston \ea~(1994)).  At intermediate stages in an accretion event, however, a
satellite would be more noticeable (\eg the Sagittarius dwarf galaxy (Ibata
\ea~1994)) and streams and moving groups can persist for more than a gigayear
(Johnston, Spergel, \& Hernquist~1995). Of course, the counter-streaming
material left by a retrograde encounter would be easily distinguished, as would
something resembling the counterrotating cores studied by Balcells \& Quinn
(1990).

\section{Discussion}

\subsection{Is our puffed up disk a ``thick disk''?}

The Milky Way's thick disk is observed to be about twice as thick as our
model's at the solar circle. Morrison (1993) quotes for the (metal-strong)
thick disk a scaleheight of about 1~kpc, $\sigma_z \sim 40$~km/s, and an
asymmetric drift of about 30~km/s.  The velocity ellipsoid is observed to be
$(\sigma_R, \sigma_\phi, \sigma_z) = (63\pm7, 42\pm4, 38\pm4)$~km/s by
Beers \& Sommer-Larsen (1995).  Our simulated disk has a smaller asymmetric
drift although a more eccentric satellite orbit might induce more lag.
Its dispersions are $(48, 42, 38)$~km/s, plus or minus a few kilometres per
second, a bit cool radially but otherwise in good agreement.  It is important
not to overinterpret this comparison, however.  The exact values we obtain
depend sensitively on the choice of scaling (\T{params}).  Moreover, our
simulation starts with a fully formed galaxy and thickens its entire disk,
whereas the Milky Way has a low-mass thick disk with an embedded high-mass
thin disk.  Thus the thick disks in the model and Milky Way are not strictly
dynamically equivalent and so not directly comparable.  The Milky Way's thick
disk is, after all, ancient, so while simulations
of this sort help us study particular processes with few competing effects,
honest comparison with observations ultimately requires that they be studied
in the context of the formation and evolution of the galaxy as a whole.

The discreteness of the Milky Way's thick disk with respect to the thin disk
argues for at least one significant ancient merger event.  A tantalizing hint
for another is the suggested substructure in the age--velocity~dispersion
relation discussed by Freeman (1993), analyzing a figure from Edvardsson
\ea~(1993).  Freeman proposes that the relation contains these three domains:
stars younger than 3~Gyr with $\sigma_z \sim 10$~km/s, stars between 3 and
10~Gyr with $\sigma_z\sim 20$~km/s, and stars older than 10 Gyr with
$\sigma_z \sim 40$~km/s (the thick disk).  Whether the transition between the
first two domains is really discrete is not clear but the figure (Figure~2 in
Freeman (1993), Figure~16b in Edvardsson \ea~(1993)) at least hints at the
possibility of a second, smaller, accretion event about 3~Gyr ago.

The thickness of our simulated disk increases at large radii because of the
inclined satellite orbit but is essentially uniform near the solar circle.
At late times, however, the bar's vertical instability causes additional
thickening (to $\langle |z| \rangle \sim 1.1$~kpc) at radii around 1
scalelength. Interestingly, an increase in the Milky Way's thick disk
scaleheight inside the solar circle is one interpretation Morrison (1993)
discusses for her data, though it is not strongly supported.  Any detection of
scaleheight varying with radius is likely to come from external galaxies, but
observers face the obstacle (which we do not) of having to subtract thin disk
and bulge profiles before searching for thick disks in noisy residuals.  In
most
instances, the issue is whether or not a thick disk has been detected, not what
its detailed structure is, so the data is generally fit with a single thick
disk
scaleheight (van Dokkum \ea~1994; Morrison, Boroson, \& Harding~1994).  The new
generation of large telescopes and large CCDs is expected to bring the study of
thick disk profiles within reach.

Given the number of parameters available for twiddling (\eg satellite mass,
density, orbital inclination, and orbital eccentricity) it seems plausible that
we could arrange a suitable thick disk by satellite accretion.  Of course,
there
is no thin disk left, a result which motivates the study of models which
include
gas (Mihos \& Hernquist 1994b; Hernquist \& Mihos 1995).  The satellite can
puff up both the gas and the existing stellar component but the gas can radiate
away its energy and resettle to form a new thin disk, leaving a thick disk
composed of a necessarily older stellar population.  The degree to which the
thick and thin disks would be distinct depends on the timescale for resettling
and the ability of the gas to form stars as it resettles.

\subsection{Is satellite accretion a good way to form bulges?}

Certainly small nuclear concentrations can be produced if the satellite core
reaches the centre.  As pointed out in \S~\ref{sect: satrem}, Hickson~87a
sports
such a feature.  There will also be a contribution from that part of the gas
which is driven to the centre (Mihos \& Hernquist 1994b; Hernquist \&
Mihos~1995).  It is not so clear that anything extended could be produced.  A
more extended satellite would simply be stripped.  A much more massive or dense
satellite would be too destructive for the disk.  The bar's vertical
instability
kicks disk material up out of the disk plane to produce a modest extended bulge
(Mihos \ea~1995), though depending on viewing angle it may be a peculiar one
(boxy, peanut-, or X-shaped).  That our initially bulgeless simulation ends up
looking very much like Hickson~87a (an S0~pec) is encouraging but gas inflow
may interfere with the bar process because steepening of the rotation
curve tends to dissipate bars (Hasan \& Norman~1990).  Even without gas inflow,
a more bar-stable model might have quickly returned to axisymmetry instead of
retaining a bar.  The exact significance of this mechanism is thus unclear but
its effectiveness here is suggestive.  In any case, it seems unlikely that it
could produce a large bulge like that of, say, NGC~5866 (Sandage \&
Bedke~1994).

Quinn \ea~(1993) noted that subsequent accretions by already thickened disks
produced much smaller changes in thickness and dispersion than the initial
accretion, so perhaps a bulge could be built up piecewise without damaging the
disk excessively.  Presumably this process would thicken any thin disk that had
reformed after the previous encounter and the gas would gradually be depleted
as some portion of it was driven to the centre each time.  This leads to a
picture in which a large disk galaxy, if it suffered several mergers with
companions of non-negligible mass, might represent a nightmare scenario for
anyone attempting to sort out its evolutionary history.  It would have a thick
disk composed of several populations of disk stars and stripped satellite
stars.
The disk would be relatively gas-poor with little present-day star formation
and relatively hot with little present-day spiral structure.  It would also
likely have a dense nucleus composed of populations from several satellites
and star formation episodes.  S0 galaxies are defined by their lack of spiral
structure and they typically also have thicker disks, larger bulges, and less
star formation than other disk galaxies (Sandage~1961; Sandage \& Bedke~1994).
Sa galaxies are defined by their tightly wrapped spiral structure, like that
visible in the last frame of \F{snaps}.  Thus mergers seem to help in all
respects to push a spiral galaxy along the Hubble sequence toward Sa/S0,
a contribution to galaxy evolution analogous to, but suitably scaled down
from, the process of two spiral galaxies merging to form an elliptical.

\subsection{Is there a constraint on cosmology?}

Unfortunately, our results do not say much about the cosmological implications
of companion accretion beyond the fact that disks are rather fragile.
Because our satellites start at the edge of the disk (to save CPU time and beat
the growth of noise), we do not really address sinking times from a
cosmological
perspective.  We find very short sinking times but a satellite at some distance
out in the halo would slowly orbit many times before reaching the disk
(\Fdot{halo}).  That is to say, the satellite is ``accreted'' by the halo long
before it is accreted by the disk, which tells us that accretion rates derived
from cosmology need to be interpreted carefully if they are to be constrained
by disk structure arguments.

\F{halo} gives sinking times as a function of satellite mass and initial
orbital
radius for circular orbits decaying by dynamical friction in an isothermal halo
(Binney \& Tremaine~1987).  At a glance one can see which kinds of objects will
hit the disk within a Hubble time or, more importantly, within the age of the
disk. Note that, while the full ranges of masses and orbital radii are
realistic
for satellite galaxies, only a small region of the diagram has any relevance:
not just any satellite can participate in the processes discussed in this
paper.
Distant, low mass satellites take longer than the age of the disk to spiral in.
The calculation is not applicable when the satellite is in the disk and any
satellite there now must have started in the shaded region.  Satellites whose
mass is comparable to the disk's mass do not qualify as ``minor'' mergers.
Evidently there is a roughly triangular region in
$R_{\mbox{init}}$~versus~$M_{\mbox{sat}}$ which we might call the ``Zone of
Interesting Parameters'', where satellites must originate to be important.
(There must also be a density limit below which a satellite will be disrupted
before reaching the disk.)  For illustration, the zone is bounded here by a
10~Gyr disk age, a 20~kpc disk radius, and a $0.25M_{\mbox{disk}}$ mass
threshold for severe destruction, but naturally the zone is much more nebulous.
In any case, the sinking times are merely rough estimates which will be
improved
only with better knowledge of the outer regions of galactic halos.  Good
galaxy-formation/cosmology simulations would be useful in this regard and would
also help us determine the expected space, energy, and angular momentum
distributions of satellites around large galaxies.

Specific comparison with the work of T\'{o}th \& Ostriker (1992) is very
difficult.  In a vague way, the fact that the damage done to our disk is
significant strengthens their result.  However, their results are averaged over
all possible orbital inclinations, making it impossible to directly relate our
numbers to theirs.  This averaging does not account for the fact that prograde
orbits are more likely to merge within a Hubble time, nor does it account for
the tendency of these orbits to settle to low inclinations, depositing most of
the vertical energy at large radii and favouring $i \sim 0$ at smaller radii.
Another difficulty is that their analysis assumes the satellite's energy is
deposited locally and its orbit decays slowly.  These assumptions might be
reasonably accurate for retrograde orbits but Figures~\ref{snaps}
and~\ref{orbit}, which illustrate the global response of the disk and the rapid
orbital decay respectively, show that our prograde mergers violate them.

T\'{o}th \& Ostriker set a limit on accretion by starting with an absolutely
cold disk and heating it with giant molecular clouds, which gives $Q=0.90$ and
a scaleheight $h=218$~pc.  They then attribute the difference between these
values and the present thin disk values to accretion. The assumptions that
total
heating and thickening scale with total accreted mass and that they are the
same
regardless of the initial heat and thickness are not strictly correct.  Quinn
\ea~(1993) found that a second satellite impact on an already heated disk had a
much less dramatic effect than the first: two satellites of 0.05$M_{disk}$ do
not equal one satellite of 0.1$M_{disk}$.  In fact, in the limit where the
satellite merger is finely subdivided into smaller events, this goes over to an
adiabatic process of slow, continuous accretion or is shut off altogether if
the objects are too distant to sink within a Hubble time (\Fdot{halo}).  Also
the assumption of a low pre-merger $Q$ means that the disk is unstable to
global
structure formation and will experience additional ``initial'' (non-merger)
heating.  Finally, when T\'{o}th \& Ostriker take account of mass loss by the
satellite (which they otherwise treat as rigid), they use the density ratio
approximation, which we have found to be inadequate (\S~\ref{sect: satrem}).
Some of these assumptions and approximations actually lead T\'{o}th \& Ostriker
to underestimate their damage; the implications of others are less clear.
Overall, it is not clear that they have succeeded in placing a limit on
accretion, nor that our results strengthen any of their conclusions.

The bottom line is that two things are needed for a limit on $\Omega_0$: a
limit
on accretion and a connection between accretion and cosmology. It is not
certain
that we have either, yet.  Meanwhile, it is worth noting that if gas can reform
a thin disk after an encounter, then the presence of a thin disk does not
necessarily mean there have been no encounters.

\section{Conclusions}

Our fully self-consistent simulations of the accretion of a companion galaxy by
a large disk galaxy allow us to examine the full interaction between the
satellite, disk, and dark matter halo.  We find that 500\,000 particles is
sufficient to eliminate numerical relaxation over the timescale of the
prograde merger, but mergers not strongly coupled to the disk (retrograde,
polar) still face competition from noise.

That the disk should dominate the interaction could have been predicted from
the
coupling between the satellite's orbit and the orbits of disk stars.  The
strong
global response of the disk depends on a satellite orbit which is both prograde
and circular.  Either high inclination or high eccentricity would curtail the
resonance considerably.  It remains to be seen whether the ``typical''
satellite
orbit becomes circular by the time it reaches the optical disk, but our initial
conditions ought to be viewed as a somewhat special case.  Future stellar
dynamical studies of satellite accretion should examine retrograde and
eccentric
orbits.  We expect them to find that the halo contributes a larger fraction of
the torque and that less of the satellite survives to the galactic centre.

With its high density and the rapid decay of its prograde orbit, our satellite
easily survives the encounter.  Half of its mass arrives in the central few
kiloparsecs of the disk, causing significant heating there.  After the merger,
the satellite core is a small, bulge-like entity in the galactic centre while
the galactic disk is rather like the thick disk component of the Milky Way.
Our particular encounter produced only about 60\% thickening at the solar
circle but enough heating so that the final
$(\sigma_R, \sigma_\phi, \sigma_z)_{\odot} \simeq (48, 42, 38)$~km/s, very
close
to the observed values for $\sigma_\phi$ and $\sigma_z$.  Our $\sigma_R$ and
asymmetric drift are lower than those observed but a more eccentric satellite
orbit might boost them a bit.  Eventually, these effects all leave the galaxy
looking like an earlier Hubble type (Sa in this case, since there is spiral
structure), supporting the picture of merger-induced galaxy evolution.

\acknowledgements

We would like to thank Heather Morrison for helpful discussions,
George Blumenthal, Caryl Gronwall, Kathryn Johnston, and Kim Supulver
for critical readings of the manuscript, and the referee, James Binney.
This work was supported in part by the San Diego Supercomputing Center,
the Pittsburgh Supercomputing Center, the Alfred P. Sloan Foundation,
NASA Theory Grant NAGW--2422, the NSF under Grants AST 90--18526
and ASC 93-18185, and the Presidential Faculty Fellows Program.

\begin{figure}
\caption{Initial radial disk structure.  Particles were evenly binned into 100
   cylindrical shells and quantities are thus averaged over azimuth.  Clockwise
   from the upper left, the panels display surface density, disk thickness,
   rotation speed and asymmetric drift, and velocity ellipsoid.  Disk thickness
   is represented by $\langle |z| \rangle$ which is related to $z_0$
   (where $z_0$ is approximately twice the exponential scaleheight; see
   eq.~\protect\ref{eq: d}) by $\langle |z| \rangle =  z_0\ln 2$, but only
   to the extent that this transformation is applicable.  Scaleheights are only
   meaningful insofar as the particle distribution fits the relevant functional
   form.  $\langle |z| \rangle$ is unambiguous because no quality-of-fit
   indicator is needed for its interpretation.
   }
\label{init}
\end{figure}

\begin{figure}
\caption{Early equilibrium disk structure.  This figure shows for an isolated
   galaxy the same quantities as \F{init} but at time $t=63$~Myr, after the
   galaxy has come into equilibrium.  For comparison, the $t=0$ structure is
   superposed with dotted curves.  Note in particular the 20\% drop in the
   central velocity dispersions.}
\label{initcomp}
\end{figure}

\begin{figure}
\caption{Structure evolution at the solar radius in isolated galaxy models.
   These are plotted on the same vertical scales as the corresponding panels
   in Figures~\protect\ref{init} and \protect\ref{initcomp} to facilitate
   visual comparison.  The solid curves represent the 450\,000-particle
   simulation, dotted represent $N=225\,000$, dot-dashed $N=90\,000$, and
   dashed $N=45\,000$.  This figure shows that we have defeated numerical
   relaxation noise over the gigayear timescale required for our prograde
   mergers.  Note that there is an initial transient because the disk does not
   start in perfect equilibrium (as shown in \Fdot{initcomp} and discussed in
   \S~2.1).  Once they have settled, the large runs remain essentially
   unchanged until their dispersions begin to creep up at $t \sim 3$~Gyr.}
\label{isostruct}
\end{figure}

\begin{figure}
\caption{Growth of $m=2$ (bar) mode in isolated galaxy models.  This figure
   displays the growth of bars in the same simulations shown in \F{isostruct}
   and also in a run with 45\,000 disk particles and a rigid halo.  The bars
set
   in at later times in larger simulations because they are seeded by random
   clumping, the amplitude of which decreases when more particles are used.
   That the halo is the source of these seeds in the self-consistent
simulations
   is made clear by the lack of growth in the rigid halo case.  The
   largest run suppresses the bar long enough for our prograde mergers to
   complete, but not our retrograde mergers.  This illustrates the necessity of
   using very large $N$, particularly in the halo, to smooth out the potential
   in self-consistent simulations.}
\label{isomodes}
\end{figure}

\begin{figure}
\caption{Face-on and edge-on views of the disk and satellite particles at equal
         intervals of $\sim 125$~Myr, starting at $t=0$.  The disk's global
         response to the satellite is quite apparent in the face-on panels
         while the thickening and
         warping of the disk are apparent in the edge-on view.  The global
         tilt is removed before further analysis by a rotation which aligns the
         total angular momentum vector of the disk particles with the $z$-axis.
         The satellite core arrives at the centre in the penultimate frame.}
\label{snaps}
\end{figure}

\begin{figure}
\caption{Satellite orbit. Clockwise from the upper left, the panels show the
   {\em cylindrical\/} radius versus time, $y$~versus~$x$ (corresponding to the
   ``face-on'' view in \F{snaps}), altitude versus radius, and altitude versus
   time.  Our satellite completes fewer than two orbits before intersecting
   the solar circle and our merger is all over by $t=1$~Gyr.  Note in
particular
   that the satellite settles into a low-inclination orbit while it is still at
   a large radius. (For comparison, the dotted lines in the lower right panel
   show the initial inclination, $i=30\arcdeg$.)}
\label{orbit}
\end{figure}

\begin{figure}
\caption{The upper panel shows the torques acting on the satellite core due to
   the disk and halo (in simulation units).  The disk is responsible for about
   75\% of the total torque integrated over the duration of the merger.
   The lower panel superposes the disk torque on the bar mode and orbital decay
   curves to make their close relationship clear.}
\label{torques}
\end{figure}

\begin{figure}
\caption{The upper panel compares the decay of the satellite orbit in four
   simulations of the same orbit (our main, prograde encounter) covering an
   order of magnitude in size to test convergence.  The consistency is very
   encouraging.  Also shown is the decay of the same orbit when a rigid halo is
   used.  The qualitative behaviour is the same but, because the satellite has
   nothing to interact with when it is far from the disk plane, the interval
   preceding the knee in the curve is longer.  Once the satellite has settled
   into the disk plane, the decay rates in the rapid sinking phase are about
the
   same because the disk dominates.  This is also true of the Quinn \ea~(1993)
   simulations except that the truncation of their disk makes the interval
   before rapid sinking even longer, about 2~Gyr.  The lower panel compares
   the decay of our fiducial orbit with that of other orbits: retrograde with
   30\arcdeg\ inclination, polar, and coplanar.  Clearly, the polar case is not
   intermediate between the prograde and retrograde cases but is essentially
   retrograde.  This demonstrates the significance of the strong disk coupling
   for prograde orbits:  it leads to a factor of three reduction in the sinking
   time (note the different scaling of the time axes in the upper and lower
   panels).}
\label{decay}
\end{figure}

\begin{figure}
\caption{Post-merger disk structure.  The solid curves show the disk in our
   encounter and the dashed curves show the disk in the isolated galaxy.
   (This comparison accomodates both the initial transient and any common
   numerical relaxation effects.)  Both are shown at $t = 1.19$~Gyr, which
   leaves time for things to settle down, although not much changes between
   $t = 1.0$~Gyr and $t = 1.2$~Gyr.  After this time, the only significant
   changes are associated with the bar's vertical instability
   (see \S~\protect\ref{sect: results}).  Note the greater radial extent
   of the disk which underwent the merger (solid curves), indicating
   conversion of satellite orbital energy to disk potential energy.}
\label{structure}
\end{figure}

\begin{figure}
\caption{Disk structure evolution at the solar circle ($R_{\odot}=8.0$~kpc).
   The top panel illustrates the thickening of the disk while the centre panel
   shows the velocity ellipsoid and asymmetric drift.  The lower panel repeats
   the orbital decay curve from \F{orbit} so that features in the disk
structure
   can be temporally matched up with the satellite's location.  In particular,
   the dotted, vertical line shows the time when the satellite crosses the
solar
   circle.  Note that most quantities rise abruptly when the satellite crosses
   $R_{\odot}$.}
\label{solar}
\end{figure}

\begin{figure}
\caption{Integrated mass distribution and surface density of the satellite
   remnant.  That almost half the satellite escapes tidal stripping and forms
   a compact central element is apparent in both panels.  The upper panel also
   shows the retrograde case (in which the satellite is nearly disrupted) for
   comparison.  The time is 1.19~Gyr.  In the lower panel, the satellite's
   surface density curve reveals that the core has not quite finished sloshing
   around the centre of the disk.  The curve for the disk and satellite
combined
   shows that the satellite core is significant enough to noticeably enhance
   the central brightness of the disk.}
\label{satstruct}
\end{figure}

\begin{figure}
\caption{Structure of disk + satellite combined.  The solid curves show the
   mass-weighted average structure of the luminous material (disk+satellite).
   For comparison, the disk is shown with dashed curves and the satellite
   remnant with dotted curves.  The time is 1.19~Gyr.  Only in the inner two
   kiloparsecs is there a sufficient density of satellite material to cause the
   combined structure to deviate substantially from the disk structure.
Outside
   the core, the satellite remnant is like a hot, thick, flared disk, although
   most of its ``thickness'' at large radii ($R>15$~kpc) actually represents
   tilting because that material is still in the original, inclined orbital
   plane.}
\label{combstruct}
\end{figure}

\begin{figure}
\caption{Sinking isochrones for an isothermal halo (based on eq.~7-27 in Binney
   \& Tremaine (1987)).  This figure provides estimates of the sinking time for
   a satellite of given mass starting in a circular orbit of given radius.
   The only objects which are significant for the present study are those which
   originate in the shaded, triangular region and are dense enough to survive
   disruption and reach the disk.  All other satellites either sink too slowly
   or are already in the disk or are so massive they will destroy the disk.
   There may also be an upper cutoff if the tidal radius of a typical halo is
   much smaller than 100~kpc.  For reference, our satellites all originate at
   the position marked with the cross, just above the lower boundary (for
   computational expedience).  That the diagram seems to predict the right
   sinking time for the prograde encounter is just an accident.  All our
   simulations started at the cross but their sinking times spanned 0.6~Gyr to
   3+~Gyr; the uncertainties in the isochrone positions are also quite large.}
\label{halo}
\end{figure}


\begin{references}

\R{} Arp, H. 1966, \apjs, 14, 1
\R{} Bahcall, J. N., Kirhakos, S., \& Schneider, D. P. 1995, \apj, 447, L1
\R{} Bahcall, J. N., Schmidt, M., \& Soneira, R. M. 1983, \apj, 265, 730
\R{} Balcells, M. \& Quinn, P. J. 1990, \apj, 361, 381
\R{} Barnes, J. E. 1988, \apj, 331, 699
\R{} Barnes, J. E. 1992, \apj, 393, 484
\R{} Barnes, J. E. \& Hernquist, L. 1992, \araa, 30, 705
\R{} Barnes, J. E. \& Hernquist, L. 1995, \apj, in press
\R{} Barnes, J. E. \& Hut, P. 1986, Nature, 324, 446
\R{} Beers, T. C. \& Sommer-Larsen, J. 1995, \apjs, 96, 175
\R{} Binney, J. J. 1992, \araa, 30, 51
\R{} Binney, J. J. \& Tremaine, S. 1987, Galactic Dynamics (Princeton:
Princeton University Press)
\R{} Burstein, D. 1979, \apj, 234, 829
\R{} Edvardsson, B., Andersen, J., Gustafsson, B., Lambert, D. L., Nissen, P.
E., \& Tomkin, J. 1993, \aap, 275, 101
\R{} Freeman, K. C. 1993, in Galaxy Evolution:  The Milky Way Perspective, ed.
Majewski, S., ASP Conference Series 49, 125
\R{} Gilmore, G., \& Reid, I. N. 1983, \mnras, 202, 1025
\R{} Gilmore, G., Wyse, R. F. G., \& Kuijken, K. 1989, \araa, 27, 555
\R{} Hasan, H. \& Norman, C. 1990, \apj, 361, 69
\R{} Hernquist, L. 1987, \apjs, 64, 715
\R{} Hernquist, L. 1989, Nature, 340, 687
\R{} Hernquist, L. 1990a, \apj, 356, 359
\R{} Hernquist, L. 1990b, J. Comput. Phys., 87, 137
\R{} Hernquist, L. 1992, \apj, 400, 460
\R{} Hernquist, L. 1993a, \apj, 409, 548
\R{} Hernquist, L. 1993b, \apjs, 86, 389
\R{} Hernquist, L. \& Mihos, J. C. 1995, \apj, 448, 41
\R{} Holmberg, E. 1941, \apj, 94, 385
\R{} Ibata, R. A., Gilmore, G. F., \& Irwin, M. J. 1994, Nature, 370, 194
\R{} Johnston, K. V., Spergel, D.N., \& Hernquist, L. 1995, \apj, 451, 598
\R{} Kuijken, K., \& Dubinski, J. 1995, \mnras, in press
\R{} Mihos, J. C. \& Hernquist, L. 1994a, \apjl, 431, L9
\R{} Mihos, J. C. \& Hernquist, L. 1994b, \apjl, 425, L13
\R{} Mihos, J. C., Walker, I. R., Hernquist, L., Mendes de Oliveira, C., \&
Bolte, M. 1995, \apjl, 447, L87
\R{} Morrison, H. L. 1993, \aj, 105, 539
\R{} Morrison, H. L., Boroson, T. A., \& Harding, P. 1994, \aj, 108, 1191
\R{} Negroponte, J. \& White, S. D. M. 1983, \mnras, 205, 1009
\R{} Pfleiderer, J. 1963, \zap, 58, 12
\R{} Pfleiderer, J. \& Siedentopf, H. 1961, \zap, 51, 201
\R{} Preston, G. W., Beers, T. C., \& Shectman, S. A. 1994, \aj, 108, 538
\R{} Quinn, P. J., \& Goodman, J. 1986, \apj, 309, 472
\R{} Quinn, P. J., Hernquist, L., \& Fullagar, D. P. 1993, \apj, 403, 74
\R{} Raha, N., Sellwood, J. A., James, R. A., \& Kahn, F. D. 1991, Nature, 352,
411
\R{} Sandage, A. 1961, The Hubble Atlas of Galaxies (Washington: Carnegie
Institution of Washington)
\R{} Sandage, A. \& Bedke, J. 1994, The Carnegie Atlas of Galaxies (Washington:
Carnegie Institution of Washington)
\R{} Schweizer, F. 1990, in Dynamics and Interactions of Galaxies, ed. Wielen,
R. (Berlin: Springer-Verlag), 60
\R{} Sellwood, J. A. 1989, \mnras, 238, 115
\R{} Toomre, A. 1981, in The Structure and Evolution of Normal Galaxies, ed.
Fall, S. M. \& Lynden-Bell, D. (Cambridge: Cambridge University Press), 111
\R{} Toomre, A. \& Toomre, J. 1972, \apj, 178, 623
\R{} T\'{o}th, G., \& Ostriker, J. P. 1992, \apj, 389, 5
\R{} Tremaine, S. 1981, in The Structure and Evolution of Normal Galaxies, ed.
Fall, S. M. \& Lynden-Bell, D. (Cambridge:  Cambridge University Press), 67
\R{} van der Kruit, P. C. \& Searle, L. 1981, \aap, 95, 105
\R{} van Dokkum, P. G., Peletier, R. F., de Grijs, R., \& Balcells, M. 1994,
\aap, 286, 415
\R{} Weinberg, M. D. 1995a, preprint
\R{} Weinberg, M. D. 1995b, private communication
\R{} Wielen, R., ed. 1990, Dynamics and Interactions of Galaxies (Berlin:
Springer-Verlag)
\end{references}
\end{document}